\documentclass[reprint,
superscriptaddress,
 amsmath,amssymb,
]{revtex4-1}

\usepackage{graphicx}
\usepackage{dcolumn}
\usepackage{bm}
\usepackage{epstopdf}

\begin{document}


\title{Gap Anisotropy in Multiband Superconductors Based on Multiple Scattering Theory}

\author{Tom G. Saunderson}
\email{t.saunderson@bristol.ac.uk}
\affiliation{HH Wills Physics Laboratory, University of Bristol, Tyndall Ave, BS8 1TL, United Kingdom}
\author{James F. Annett}%
\affiliation{HH Wills Physics Laboratory, University of Bristol, 
Tyndall Ave, BS8 1TL, United Kingdom}
\author{Bal\'azs \'Ujfalussy}%
\affiliation{Institute for Solid State Physics and Optics, Wigner Research Centre for Physics, Hungarian Academy of Sciences, P.O. Box 49, H-1525 Budapest, Hungary}
\author{G\'abor Csire}%
\affiliation{Catalan Institute of Nanoscience and Nanotechnology (ICN2), CSIC, BIST, Campus UAB, Bellaterra, Barcelona, 08193, Spain}
\affiliation{Institute for Solid State Physics and Optics, Wigner Research Centre for Physics, Hungarian Academy of Sciences, P.O. Box 49, H-1525 Budapest, Hungary}
\affiliation{HH Wills Physics Laboratory, University of Bristol, Tyndall Ave, BS8 1TL, United Kingdom}
\author{Martin Gradhand}%
\affiliation{HH Wills Physics Laboratory, University of Bristol, Tyndall Ave, BS8 1TL, United Kingdom}

\date{\today}

\begin{abstract}
We implement the Bogoliubov-de Gennes (BdG) equation in a screened Korringa-Kohn-Rostoker (KKR) method for solving, self-consistently, the superconducting state for 3d crystals. This method combines the full complexity of the underlying electronic structure and Fermi surface geometry with a simple phenomenological parametrisation for the superconductivity. We apply this theoretical framework to the known s-wave superconductors Nb, Pb, and MgB$_2$. In these materials multiple distinct peaks at the gap in the density of states were observed, showing significant gap anisotropy which is in good agreement with experiment. Qualitatively, the results can be explained in terms of the k-dependent Fermi velocities on the Fermi surface sheets exploiting concepts from BCS theory. 
\end{abstract}


\pacs{Valid PACS appear here}
\maketitle


\section{\label{sec:1}Introduction}
The structure of the superconducting gap in s-wave phonon mediated superconductors may show a surprisingly complex structure. Not only can it show a large degree of anisotropy on the Fermi surface of a given material but several known examples exhibit clear signatures of two superconducting gaps. The most notable of cases are Pb \cite{Tomlinson1976,Blackford1971,Short2000,Ruby2015b} and MgB$_2$ \cite{Szabo2001,Choi2002,Mou2015} and to a lesser extent Nb \cite{Dobbs1964,MacVicar1968,Berndt1970,Klaumunzer1974,Hahn1998} with conflicting reports \cite{Anderson1971,Novotny1975}. While it is understood that the gap anisotropy arises in these systems from multiple Fermi surface sheets, not many phonon mediated s-wave superconductors showing multiple gaps have been identified. 


In the original BCS theory \cite{Bardeen1957} only a single spherical band was considered and so it was impossible to obtain any gap anisotropy. In addition, it was limited to the weak coupling regime, which later was rectified by the Eliashberg theory \cite{Eliashberg1960,Carbotte1990} and more recently by full DFT approaches to electron phonon coupling driven superconductivity \cite{Luders2005,Marques2005}. Although gap anisotropy analysis is present in such codes, much of the DFT based \textit{ab initio} work focused on other aspects. These aspects being; expanding the description of s-wave superconductivity with emphasis on the correct description of the driving mechanism \cite{Margine2013,Flores-Livas2016,Sanna2018,Schrodi2018,Gastiasoro2019}, or on the extension to unconventional pairing pushing the boundaries in our treatment of iron based \cite{Essenberger2016}, spin \cite{Essenberger2014} and magnetic effects \cite{Linscheid2015,Linscheid2015a}.


Beyond the \textit{ab initio} work on superconductivity, there is also extensive work on using parametrised models to describe superconductivity \cite{Sigrist1991,Mackenzie2003,Annett2006,Hillier2012,Mazidian2013,Singh2014,Weng2016,Robbins2017,Kreisel2017,Brydon2018}. These models are a powerful tool for describing unconventional superconductors and some of the basic principles of standard s-wave superconductors. However these models have the drawback that they require the normal state to be parametrised from either experimental data or a density functional theory (DFT) calculation. This often results in over-simplification of the normal state band structure in order to construct an efficient model to describe the superconducting state. While this usually leads to deep understanding of aspects of the superconductivity, it might obscure the importance of the complexity and orbital hybridisation of the underlying electronic structure.


In this work we aim to follow a route distinct to those two main directions. We aim to describe the full complexity of the normal state electronic structure whilst treating the superconducting pairing interaction within a simplified model. This idea is very similar to the LDA+U  method where the Hubbard U interaction is also treated as a tunable parameter \cite{Anisimov1997}. This has proven to be very successful in modelling strongly correlated materials without compromising on the full electronic structure within a strongly correlated state.

In analogy to this, our method leads to a full electronic structure within the superconducting state giving access to the full gap anisotropy in multi sheeted s-wave superconductors. We build upon previous work from some of the authors on the implementation of the scalar relativistic \cite{Csire2015a,Csire2016a,Csire2016b} and fully relativistic \cite{Csire2018,Csire2018a} BdG equations in a layered KKR DFT code. The advantage of this approach is the access to the full normal state electronic structure from first principles without using simplified models. By just considering s-wave pairing and one effective pairing parameter, we show quantitatively the complexity of the superconducting gap including its full anisotropy even in simple elemental crystals. 

The work is structured as follows. First, we will outline the basic theoretical background of the method including the differences in implementation to the earlier work \cite{Csire2015a,Csire2016a,Csire2016b}. This will be followed by numerical tests to determine the robustness and accuracy of the theoretical framework, namely the solution of the scalar relativistic BdG equations. In the next part we will present several examples of simple s-wave superconductors, the resulting gap anisotropy on the various Fermi surface sheets, and some simple arguments on how the normal state properties drive the observed anisotropies. In all cases we will relate our work to experimental observations.  

\section{\label{sec:2}Method}

This section will describe the implementation of superconductivity into the existing KKR code \citep{Gradhand2009}. The theory for the single site solver and multiple scattering terms have been introduced by G. Csire \textit{et al} \citep{Csire2015a}. All equations are given in Rydberg atomic units.

The density functional theory for superconductors was initially presented by L. N. Oliveira, E. K. U. Gross and W. Kohn \cite{Oliveira1988}, who introduced the effective pairing interaction $\Delta_{eff}(\mathbf{r},\mathbf{r}')$, describing the superconducting state in addition to the conventional Kohn-Sham potential $V_{eff}(\mathbf{r})$. These two potentials are defined as 
\begin{align}
\label{eqn:Veff}
V_{eff}(\mathbf{r}) &= V_{ext}(\mathbf{r}) + \int d^3 r  \frac{\rho(\mathbf{r})}{|\mathbf{r}-\mathbf{r}'|} + \frac{\delta E_{xc}[\rho,\chi]}{\delta \rho(\mathbf{r})}, \\
\label{eqn:Deltaeff_1}
\Delta_{eff}(\mathbf{r},\mathbf{r}') &= \frac{\delta E_{xc}[\rho,\chi]}{\delta \chi(\mathbf{r},\mathbf{r}')},
\end{align}
where $\chi(\mathbf{r},\mathbf{r}')$ is the anomalous density and $E_{xc}[\rho,\chi]$ is the exchange correlation functional for a superconductor. The exchange correlation functional was later approximated \cite{Suvasani1993} as
\begin{align}
\label{eqn:Suvasini}
E_{xc}[\rho,\chi] = &E^0_{xc}[\rho] -\int d^3 r_1 d^3 r_2 d^3 r'_1 d^3 r'_2 \\
&\chi^*(\mathbf{r}_1,\mathbf{r}'_1)\Lambda[\rho,\chi](\mathbf{r}_1,\mathbf{r}'_1,\mathbf{r}_2,\mathbf{r}'_2)\chi(\mathbf{r}_2,\mathbf{r}'_2), \nonumber
\end{align}
where $E^0_{xc}[\rho]$ is the normal state exchange-correlation functional, and $\Lambda[\rho,\chi](\mathbf{r}_1,\mathbf{r}'_1,\mathbf{r}_2,\mathbf{r}'_2)$ is the pairing kernel. Similarly to ref. \cite{Suvasani1993}, in the KKR framework we consider the atomic sphere approximation (ASA). The kernel is then approximated to
\begin{equation}
\Lambda[\rho,\chi](\mathbf{r}_1,\mathbf{r}'_1,\mathbf{r}_2,\mathbf{r}'_2) = \Lambda\delta(\mathbf{r}_1-\mathbf{r}'_1)\delta(\mathbf{r}_1-\mathbf{r}_2)\delta(\mathbf{r}_1-\mathbf{r}'_2),
\end{equation}
inside the ASA spheres and
\begin{equation}
\Lambda[\rho,\chi](\mathbf{r}_1,\mathbf{r}'_1,\mathbf{r}_2,\mathbf{r}'_2) = 0,
\end{equation}
outside, where $\Lambda$ is called the interaction parameter. A further conventional simplification is for $\Delta_{eff}(\mathbf{r},\mathbf{r}')$ and $\chi(\mathbf{r},\mathbf{r}')$ to be local, resulting in
\begin{align}
\Delta_{eff}(\mathbf{r},\mathbf{r}') &= \Delta_{eff}(\mathbf{r})\delta(\mathbf{r}-\mathbf{r}'), \\
\chi(\mathbf{r},\mathbf{r}') &= \chi(\mathbf{r})\delta(\mathbf{r}-\mathbf{r}').
\end{align}
Equation (\ref{eqn:Deltaeff_1}) therefore becomes
\begin{equation}
\Delta_{eff}(\mathbf{r}) = \Lambda \chi(\mathbf{r}).
\label{eqn:Deltaeff}
\end{equation}
When the spherical symmetry (ASA) and local approximation for the pairing potential are combined, we restrict ourselves to s-wave superconductivity. Any kind of non-s wave superconductivity would need a non-spherical pairing potential, coupling between different orbital channels or non-locality in the pairing potential. In this report $\Lambda$ is tuned such that the gap in the density of states matches experimental results for the zero temperature gap size $\Delta(T=0)$ of the material in question. Further details of this are discussed in section \ref{sec:3}. 

The KKR method exploits a local atomic basis that uses multiple scattering theory, which gives access to the full Green's function of the system. For details on the implementation we refer to G. Csire \textit{et al} \citep{Csire2015a}. Here we restrict the discussion to the essential components to highlight differences in the implementation and to shed light onto the results. The Green's function for the system is defined as
\begin{equation}
\hat{G}_{BdG}(z) = 
\left(\begin{matrix}
\hat{G}^{ee}(z) & \hat{G}^{eh}(z) \\
\hat{G}^{he}(z) & \hat{G}^{hh}(z) 
\end{matrix}\right) =
\big(z\hat{I} - \hat{H}_{BdG}\big)^{-1},
\end{equation}
where $\hat{H}_{BdG}(\mathbf{r}) = \langle \mathbf{r}|\hat{H}_{BdG}|\mathbf{r}\rangle$ and
\begin{align}
\hat{H}_{BdG}(\mathbf{r}) &= 
\left(\begin{matrix}
\hat{H}(\mathbf{r}) & \Delta_{eff}(\mathbf{r})\\
\Delta_{eff}(\mathbf{r})^* & -\hat{H}(\mathbf{r})^*
\end{matrix}\right), \\
\hat{H}(\mathbf{r}) &= -\nabla^2 + V_{eff}(\mathbf{r}) - \mu.
\end{align}
Here, $\mu$ is the chemical potential and $z = \epsilon + i\delta$. The densities $\rho(\mathbf{r})$ and $\chi(\mathbf{r})$ can be calculated by taking the trace of different components of $G_{BdG}(z,\mathbf{r},\mathbf{r}')$,
\begin{align}
\label{eqn:rho}
\rho(\mathbf{r}) = &-\frac{1}{\pi}\int^{\infty}_{-\infty} d\epsilon f(\epsilon) \mathrm{Im}\mathrm{Tr}G^{ee}(\epsilon,\mathbf{r},\mathbf{r}') \nonumber\\
&-\frac{1}{\pi}\int^{\infty}_{-\infty} d\epsilon [1-f(\epsilon)] \mathrm{Im}\mathrm{Tr}G^{hh}(\epsilon,\mathbf{r},\mathbf{r}'),\\
\label{eqn:chi}
\chi(\mathbf{r}) = &-\frac{1}{4\pi}\int^{\infty}_{-\infty} d\epsilon [1-2f(\epsilon)] \mathrm{Im} \mathrm{Tr} G^{eh} (\epsilon,\mathbf{r},\mathbf{r}') \nonumber \\
&-\frac{1}{4\pi}\int^{\infty}_{-\infty} d\epsilon [1-2f(\epsilon)]  \mathrm{Im} \mathrm{Tr} G^{he} (\epsilon,\mathbf{r},\mathbf{r}'),
\end{align}
where the limit is taken such that $\delta \rightarrow 0^{+}$. We use equations (\ref{eqn:rho}) and (\ref{eqn:chi}) to find new expressions for $V_{eff}(\mathbf{r})$ and $\Delta_{eff}(\mathbf{r})$ using equations (\ref{eqn:Veff}) and (\ref{eqn:Deltaeff}). From here a new $\hat{H}_{BdG}(\mathbf{r})$ is constructed to calculate a new Green's function $\hat{G}_{BdG}(\epsilon,\mathbf{r},\mathbf{r}')$, and thus self consistency can be achieved. 

The ASA approximation sets a boundary to each atomic site, $i$, called the ASA radius $r^{ASA}_i$. This approximation implies that the potentials $V_{eff}(\mathbf{r})$ and $\Delta_{eff}(\mathbf{r})$ can be written in sums
\begin{align}
V_{eff}(\mathbf{r}) &= \sum_i V_i(\mathbf{r}), \\
\Delta_{eff}(\mathbf{r}) &= \sum_i \Delta_i(\mathbf{r}),
\end{align}
and ensures that $V_i(\mathbf{r}) = 0$ and $\Delta_i(\mathbf{r}) = 0$ if $|\mathbf{r}| \geq r_i^{ASA}$. The analogue of equation (\ref{eqn:Deltaeff}) becomes, 
\begin{equation}
\Delta_i(\mathbf{r}) = \Lambda_i\chi_i(\mathbf{r}).
\end{equation}

The resulting code is therefore able to perform calculations which relax both charge and anomalous densities along with the Fermi energy. In practice, we first relax the densities and only in the last steps allow the Fermi energy to relax as well. We found that relaxing the Fermi energy does not significantly change the solution and is computationally expensive. The results we present in this report are therefore fixed Fermi level calculations.  

\section{\label{sec:3}Computational Details}



First, we test the numerical parameters and the resulting gaps within our framework.
We define the average gap \cite{Suvasani1993}
\begin{equation}
\bar{\Delta}_r = \frac{1}{V_{WS}}\int_{WS} \Delta_{eff}(\mathbf{r}) d^3r\, ,
\label{eqn:AvGap}
\end{equation}
where $V_{WS}$ represents the volume of the Wigner-Seitz cell. This is a practical expression, however, in a real bandstructure this average gap is not the same as the actual gap in the density of states.  

\begin{figure}[h]
\centering
\includegraphics*[width=1\linewidth,clip]{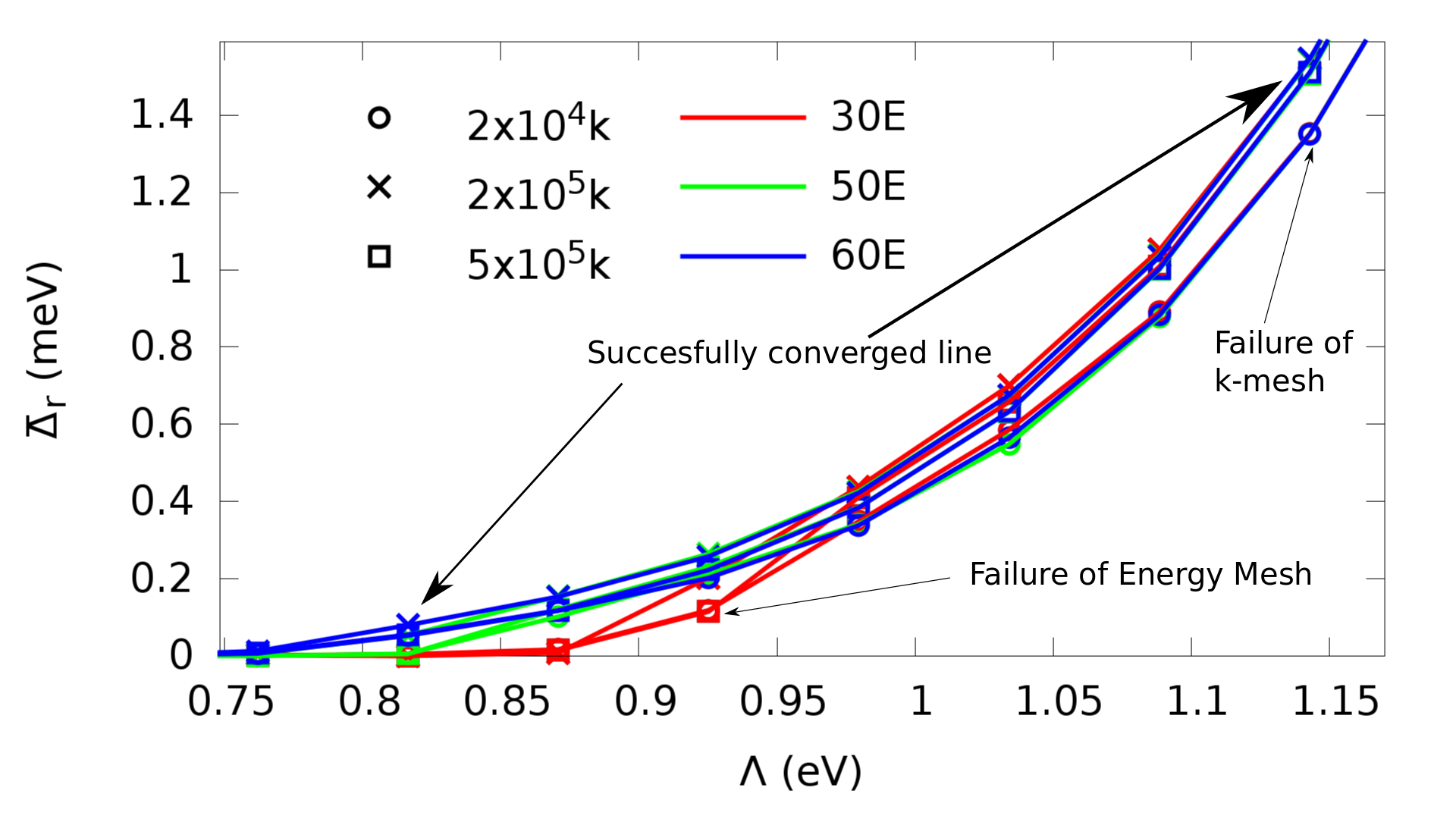}
\caption{The convergence test performed on a non-relativistic calculation of Niobium. The parameters are the energy and k-mesh. The k-mesh is the actual number of k-points used to calculate the energy points closest to the superconducting gap. Plot symbols denote $2\times10^4$ (circle), $2\times10^5$ (cross) and $5\times10^5$ (square) k-points, colours denote 30 (red), 50 (green) and 60 (blue) energy points. } 
\label{fig:Convergence}
\end{figure}

Fig.~\ref{fig:Convergence} shows the results of the convergence tests for non-relativistic (NR) Niobium. For low $\Lambda$, all calculations with 30 energy points are suppressed relative to calculations with a higher energy meshes. At high $\Lambda$ the solutions with the same sized k-mesh group together. We decided that 50 energy points and $200 \times 10^3$ k-points is the best trade off between accuracy and computational cost.

\begin{figure}[h]
\centering
\includegraphics*[width=1\linewidth,clip]{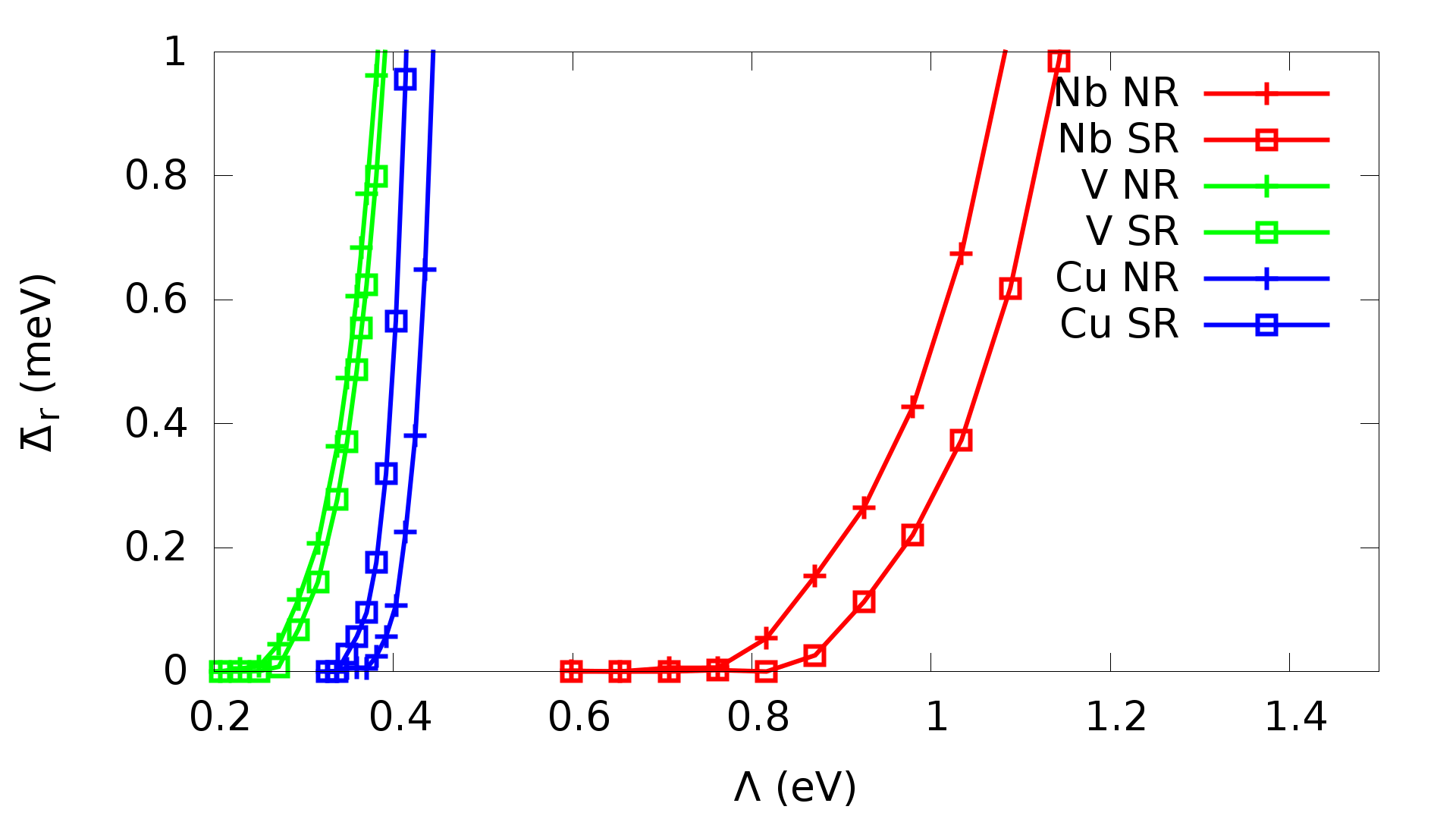}
\caption{The comparison of the average gap size of Nb, V and Cu for non relativistic and scalar relativistic calculations as a function of $\Lambda$.} 
\label{fig:SR_v_NR}
\end{figure}

For the remainder of the report, we perform calculations using the scalar-relativistic (SR) BdG equations. A comparison between NR and SR results is shown in Fig.~\ref{fig:SR_v_NR} for Nb, V and Cu. There is no clear trend in terms of the gap size going from NR to SR. For V and Nb the gap is reduced, for Cu it is increased. The influence becomes more dominant with larger atomic number as expected when including scalar relativistic corrections. 

Within this framework, $\Lambda$ is the only free parameter. In general this parameter is tuned such that the resulting gap matches the experimental zero temperature gap size. The complication, especially for anisotropic gaps or multigap systems, is what definitions are used experimentally and theoretically to establish the gap size. In the following we discuss how in practice we constrain our free parameter $\Lambda$. The first definition is using the average gap from equation (\ref{eqn:AvGap}). This quantity, however, is just the integrated anomalous density of states $\chi(\mathbf{r})$ multiplied by a constant, and does not necessarily relate to the gap seen in the density of states. It can been easily proven comparing the gaps in Fig.~\ref{fig:DOS} and $\bar{\Delta}_\mathbf{r}$ for Nb in table \ref{table:Params}. Clearly these numbers are different from each other. In fact, by comparing the values for $\bar{\Delta}_\mathbf{r}$ for Pb and Nb it is possible to see that Nb has a smaller $\bar{\Delta}_\mathbf{r}$ than Pb. As Nb has a larger $T_c$ than Pb it is obvious that the real gap size must be larger. We believe that the fundamental reason for this discrepancy lies in the fact that the relationship between the effective pairing interaction $\Delta_{eff}(\mathbf{r})$ and the k-dependent gap $\Delta(\mathbf{k})$ is not trivial. Similarly, the average gap found in the literature ($\bar{\Delta}_\mathbf{k}$) and shown in table \ref{table:Params}, column 6, is conventionally obtained from  a k-space integration of $\Delta(\mathbf{k})$. This quantity is distinct to the average $\bar{\Delta}_\mathbf{r}$ calculated from the real space integral over $\Delta_{eff}(\mathbf{r})$ (see Table \ref{table:Params}, column 5). For Nb and Pb, $\bar{\Delta}_\mathbf{r}$ is smaller than $\bar{\Delta}_\mathbf{k}$ but for MgB$_2$ it is larger.

Finally, we would like to summarize the numerical parameters used throughout the papers. After convergence of the potentials is reached, the imaginary part of the energy, $\delta = 2\mu \mathrm{Ry}$, $0.2\mu \mathrm{Ry}$, $10\mu \mathrm{Ry}$ is used for Nb, Pb and MgB$_2$ respectively. These are the parameters for the DOS and Bloch spectral function calculations when focussing on the gap structure as well as for $\Delta(\mathbf{k})$. The number of k-points in the DOS calculations are $10\times10^{6}$ for Nb and Pb and $2.7\times10^{6}$ for MgB$_2$.

\begin{table*}
\centering
\begin{tabular}{|c|c|c|c|c|c|}
\hline
\rule{0pt}{2.5ex} Element & 
$D(\epsilon_F)$ (Ry$^{-1}$) & 
$\Lambda$ (Ry) & 
$\bar{\Delta}_\mathbf{r}$ (meV) & 
$\Delta^{exp}$ (meV) & 
$\bar{\Delta}_\mathbf{k}$ (meV) \\
\hline
\hline
\rule{0pt}{2.5ex} Nb & 17.72 & 0.0846 & 1.05 & 1.56\cite{Savrasov1996},(1.79, 1.64, 1.20)\cite{Hahn1998} & 1.53\cite{Savrasov1996} \\

\rule{0pt}{2.5ex} Pb & 6.77 & 0.348 & 1.15 & 1.33\cite{Savrasov1996},(1.40, 1.27)\cite{Ruby2015b} & 1.35\cite{Savrasov1996},(1.0, 0.8)\cite{Floris2007} \\

\rule{0pt}{2.5ex} MgB$_2$ & 3.83 & 0.288 & 7.93 & (7.0, 3.0)\cite{Mou2015} & (6.8, 2.45)\cite{Floris2005} \\
\hline
\end{tabular}
\caption{$D(\epsilon_F)$ is the density of states at the Fermi level in the normal state obtained from our calculations. $\Lambda$ is the interaction parameter used in this investigation to match our calculations to the experimental zero temperature gap, more details in section \ref{sec:4}. The average gap, $\bar{\Delta}_\mathbf{r}$, is calculated using (\ref{eqn:AvGap}). $\Delta^{exp}$ are average gaps from experiments \cite{Savrasov1996,Ruby2015b,Mou2015,Hahn1998}, $\bar{\Delta}_\mathbf{k}$ are average gaps from theoretical $\Delta(\mathbf{k})$ integrations \cite{Savrasov1996,Floris2005,Floris2007}.} 
\label{table:Params}
\end{table*}

\section{\label{sec:4} Results}
\subsection{\label{sec:4a} Niobium}

\begin{figure}[h]
\centering
\includegraphics*[width=1\linewidth,clip]{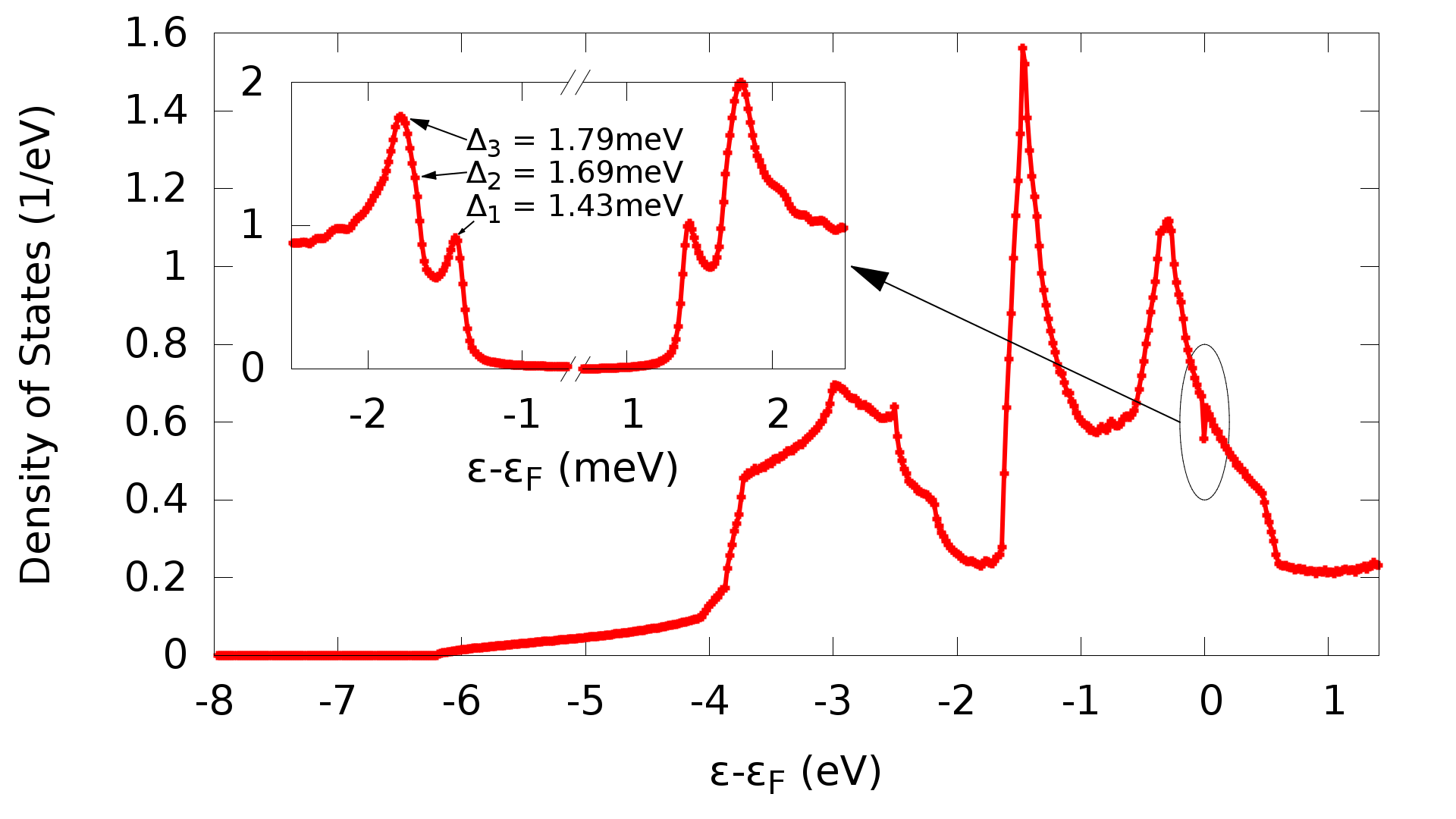}
\caption{A scalar relativistic calculation of the electronic density of states of Niobium in the superconducting state, with the inset showing the gap at the Fermi level $\epsilon_F$.} 
\label{fig:DOS}
\end{figure}

For calculations in the case of Nb, we tuned $\Lambda$ such that the gap in the density of states around the Fermi level was matched to the experimental gap size via tunnelling experiments\cite{Hahn1998}. It predicted different sizes for the superconducting gap depending on the exposed surface of the single crystalline Nb.  The different crystal planes investigated were [001], [110] and [111]. The values for the superconducting gap are given by $\Delta_{001} = 1.20$meV, $\Delta_{110} = 1.79$meV and $\Delta_{111} = 1.64$meV. We chose to tune $\Lambda$ such that the outer peak of our superconducting gap matched $\Delta_{110}$. The result is shown in Fig.~\ref{fig:DOS}. In the inset we can identify two clear peaks and one weak shoulder corresponding to three distinct gaps at $\Delta_{1} = 1.43$meV, $\Delta_{2} = 1.69$meV and $\Delta_{3} = 1.79$meV, which is in reasonable quantitative agreement to the experiment. In comparison to other literature the gap anisotropy is also well matched \cite{Dobbs1964,MacVicar1968,Berndt1970,Klaumunzer1974}.

\begin{figure}[h]
\centering
\includegraphics*[width=1\linewidth,clip]{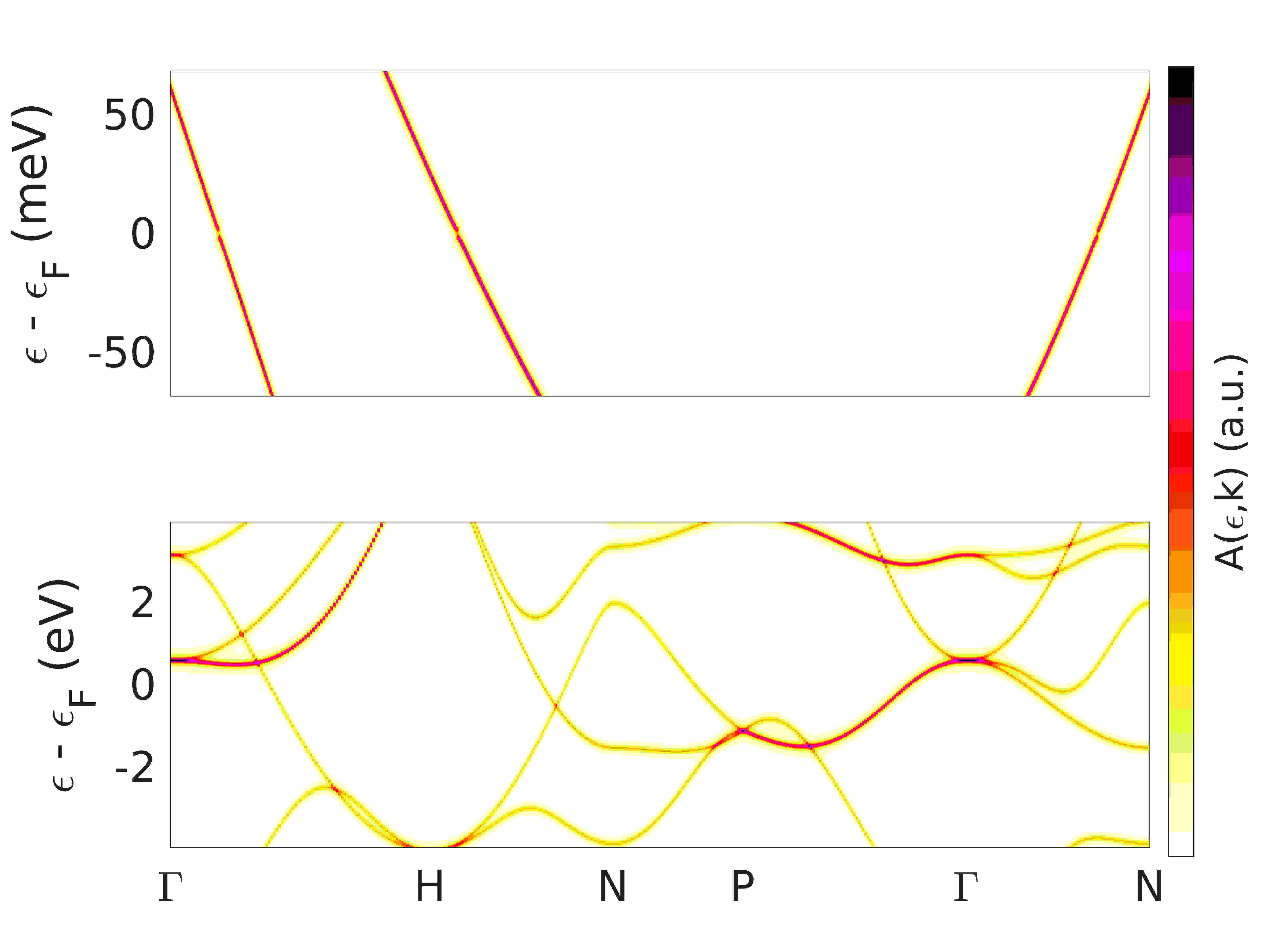}
\caption{A Bloch spectral function of Niobium in the superconducting state, showing the bandstructure in directions of high symmetry. The units of the spectral function are arbitrary. In the top panel with a higher energy resolution the superconducting gap is just about visible.} 
\label{fig:BandStruct}
\end{figure}

\begin{figure}[h]
\centering
\includegraphics*[width=1\linewidth,clip]{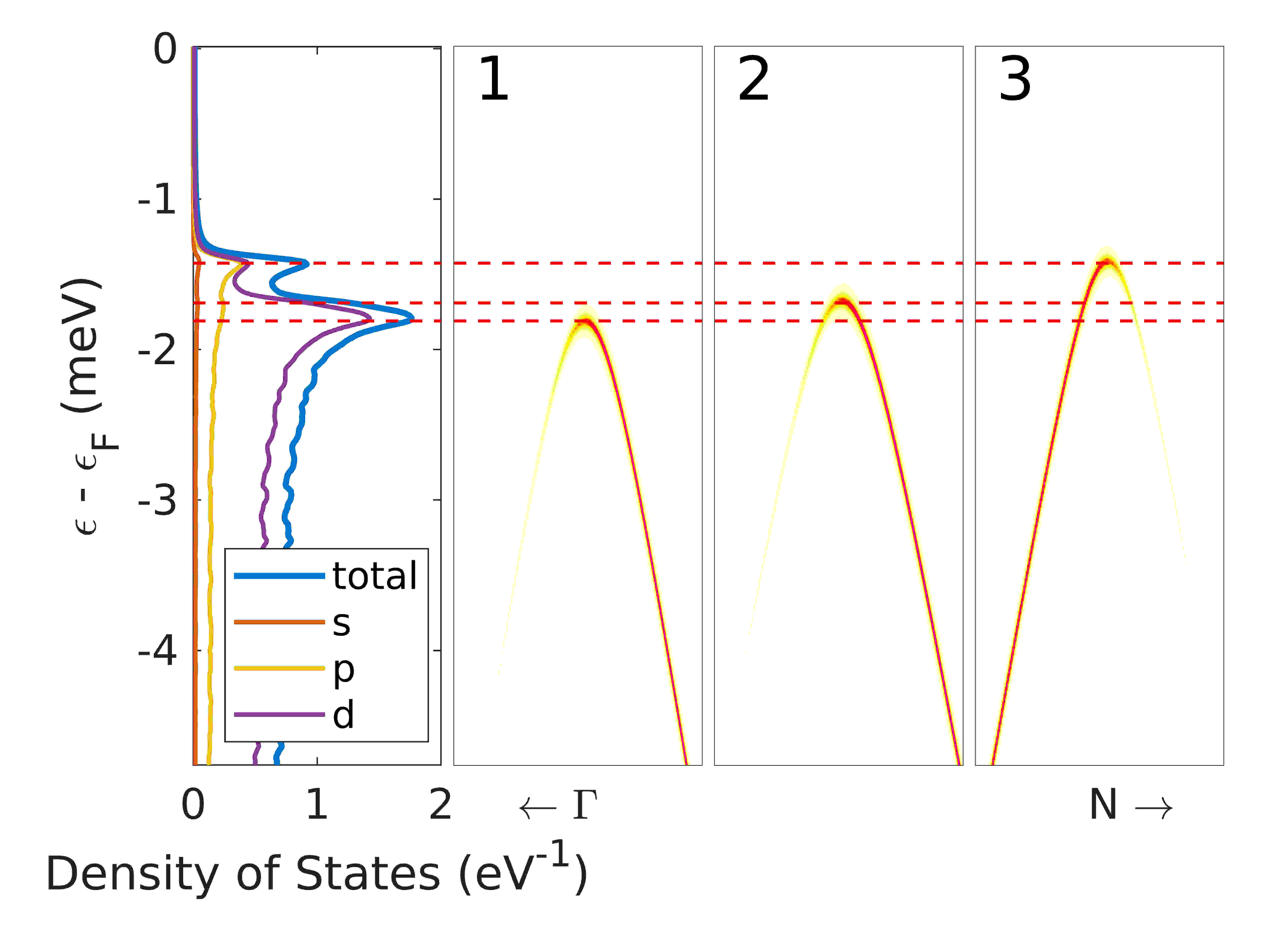}
\caption{Left Panel: The total and orbital resolved density of states of Niobium near the Fermi level. Panel 1-3: Band gaps in the $\Gamma$ to $N$ direction. Panels 1 and 2 are associated with the larger gap in the DOS, and panel 3 is associated with the smaller gap. The units of the spectral function are arbitrary.}
\label{fig:GapAnisotropy}
\end{figure}

\begin{figure}[h]
\centering
\includegraphics*[width=1\linewidth,clip]{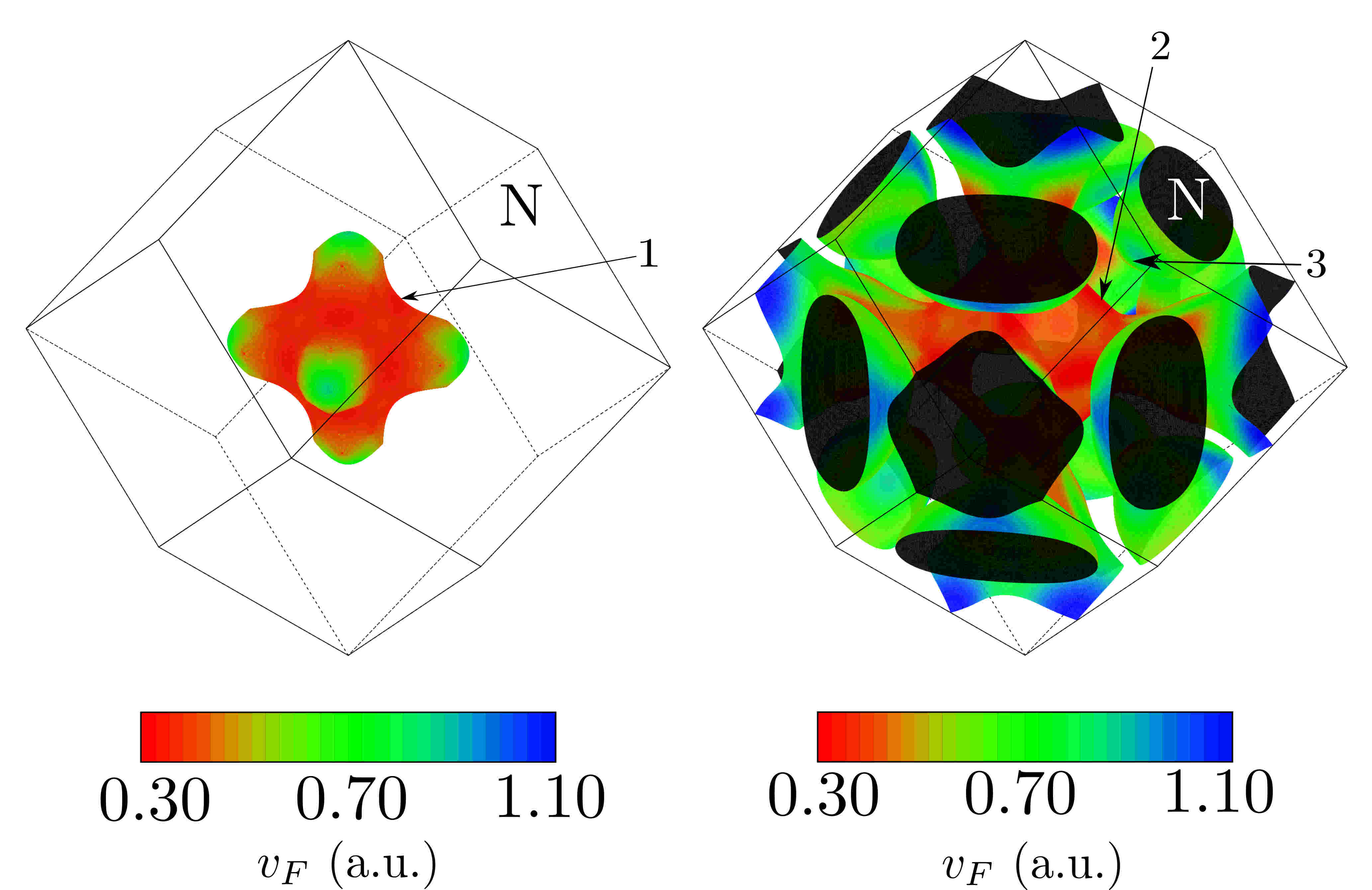}
\caption{The Fermi surface sheets with the color scale showing the Fermi velocity of Nb. The labels refer to the gaps identified in Fig.~\ref{fig:GapAnisotropy}.} 
\label{fig:FermiSurface_vf}
\end{figure}

The lower panel of Fig.~\ref{fig:BandStruct} shows the Bloch spectral function for a large energy window but in the top panel we focus on the region around the Fermi level along the $\Gamma$ to $N$ direction with a smaller energy window ($-50\mathrm{meV}$ to $50\mathrm{meV}$). On that scale the opening of the superconducting gap is just about resolved. In order to investigate the gap and the associated anisotropy as highlighted in Fig.~\ref{fig:DOS} the relevant energy resolution is $-4.5$meV to $0$meV  and for the bandstructure we focus on the high symmetry line $\Gamma$ to $N$ in Fig.~\ref{fig:GapAnisotropy}. A double peaked superconducting gap is clearly resolved, with band gaps 1 and 2 contributing to the outermost gap, and the 3rd band gap relating to the inner peak. In order to understand this effect we consider the orbital character associated with each band. While the inner peak is `p-d' hybridised, the outer peak is almost entirely of a $\mathrm{d}$-electron character. Typically, `p' character bands will show a larger Fermi velocity at the Fermi level compared to the `d' character bands. This is confirmed by showing the Fermi velocities as a colour map on the distinct Fermi surface sheets in Fig.~\ref{fig:FermiSurface_vf}, where points of the Fermi surface associated with panels 1-3 from Fig.~\ref{fig:GapAnisotropy} are labelled. Evidently panels 1 and 2 have similar velocities at the point of crossing, whereas panel 3 has a velocity which is noticeably larger. Since the DOS is inversely proportional to the Fermi velocity,
\begin{equation}
D(\epsilon_F) \propto \frac{1}{v_F},
\label{eqn:DOSvVf}
\end{equation} 
and within the BCS theory \cite{Bardeen1957,Ketterson1999a} the gap scales with the DOS
\begin{equation}
\Delta \approx 2k_B\Theta_D \exp\left(-\frac{1}{V D(\epsilon_F)}\right),
\label{eqn:BCS}
\end{equation} 
we can infer a gap anisotropy which is related to the distribution of Fermi velocities on the Fermi surface sheets.  

\begin{figure}[h]
\centering
\includegraphics*[width=1\linewidth,clip]{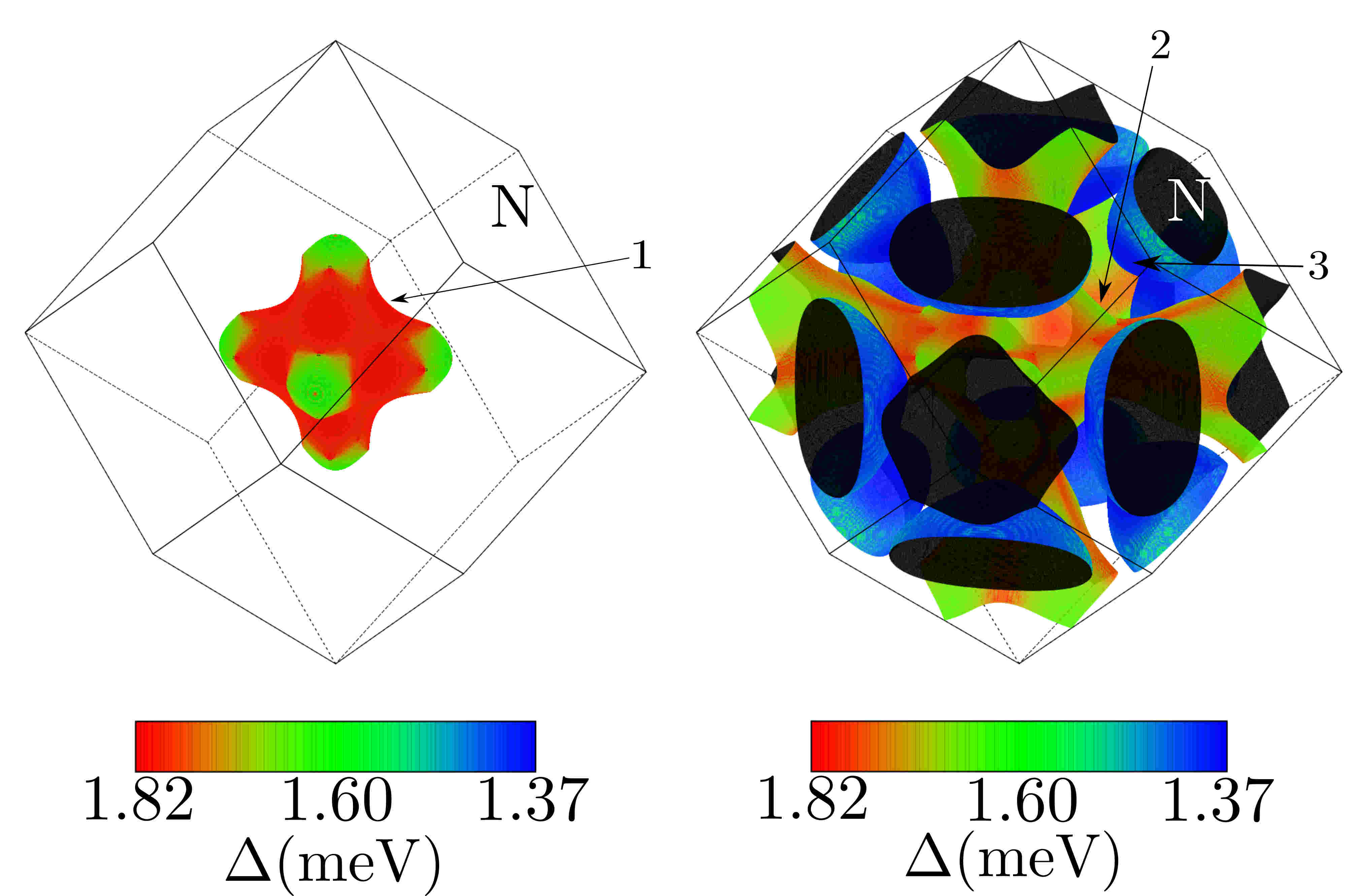}
\caption{The Fermi surface sheets of Nb in the normal state with the gap size of Nb in the superconducting state superimposed as a colour scale on top. The labels refer to the panels in Fig.~\ref{fig:GapAnisotropy} and identify the points on the Fermi surface where the gaps in Fig.~\ref{fig:GapAnisotropy} appear.} 

\label{fig:FermiSurface_gap}
\end{figure}

In Fig.~\ref{fig:FermiSurface_gap} we extend this analysis to the full Fermi surface and in comparing it to Fig.~\ref{fig:FermiSurface_vf} a strong correlation between $\mathrm{v}_F(\mathbf{k})$ and $\Delta(\mathbf{k})$ is observed. 

\subsection{\label{sec:4b} Lead }

Lead often has disordered or amorphous crystal structures. Experimentally, this has been avoided in STM experiments by M. Ruby \textit{et al}\cite{Ruby2015b}. In this experiment the authors could identify two distinct peaks of the superconducting gap at an energy separation of 150$\mathrm{\mu eV}$. Within our calculations the fcc crystal structure with a lattice constant of 4.95$\mathrm{\AA}$ is used. Here we restrict the computation to three dimensional periodic crystals but future work will focus on describing the surface explicitly as measured in the experiment. For the bulk material we assume that the gap anisotropy in Pb should at least be of similar order as found in the STM experiments. 

\begin{figure}[h]
\centering
\includegraphics*[width=1\linewidth,clip]{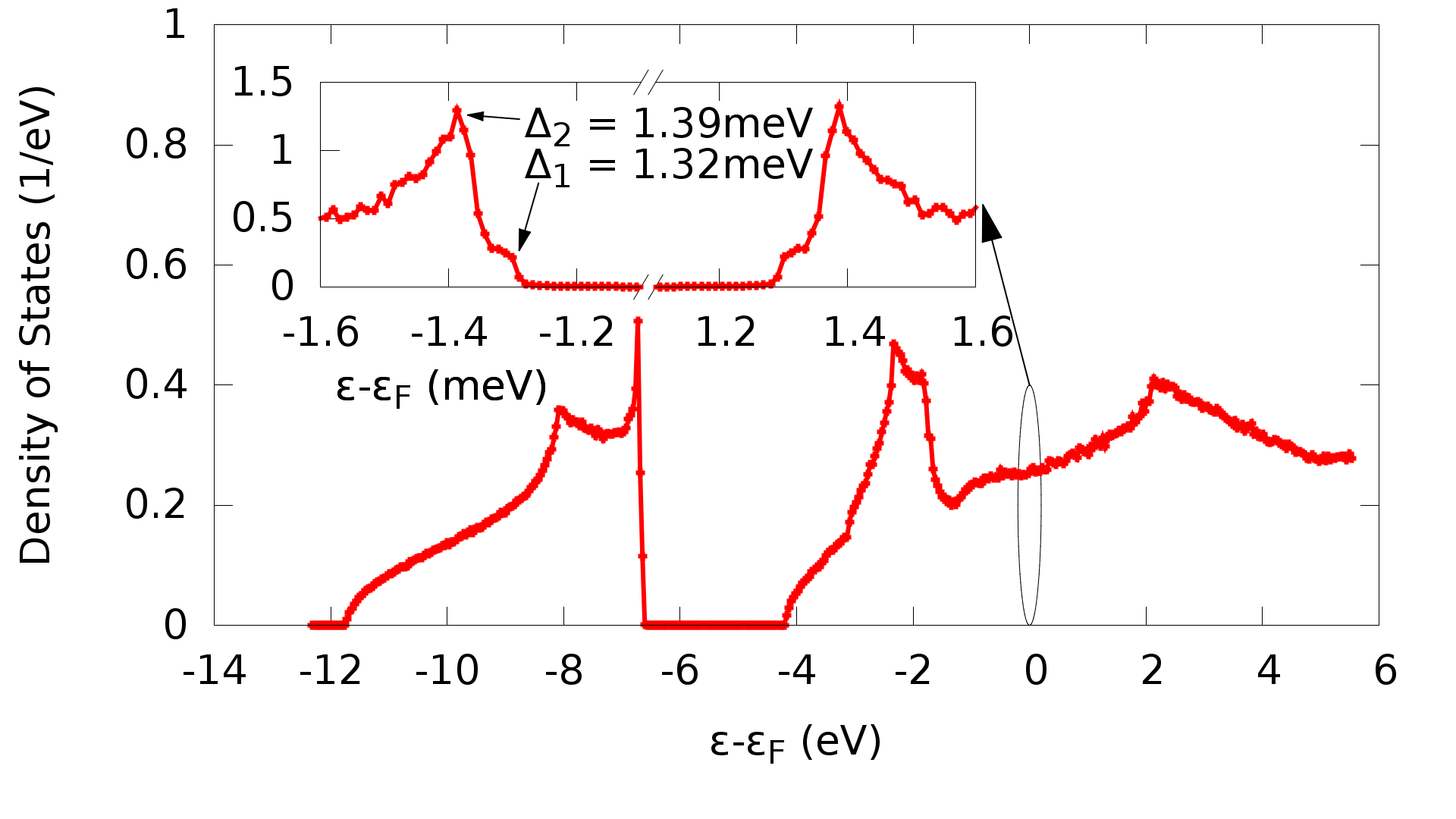}
\caption{A scalar relativistic calculation of the electronic density of states of Lead in the superconducting state with an inset figure showing the peaks of the superconducting gap.} 
\label{fig:DOSPb}
\end{figure}

\begin{figure}[h]
\centering
\includegraphics*[width=1\linewidth,clip]{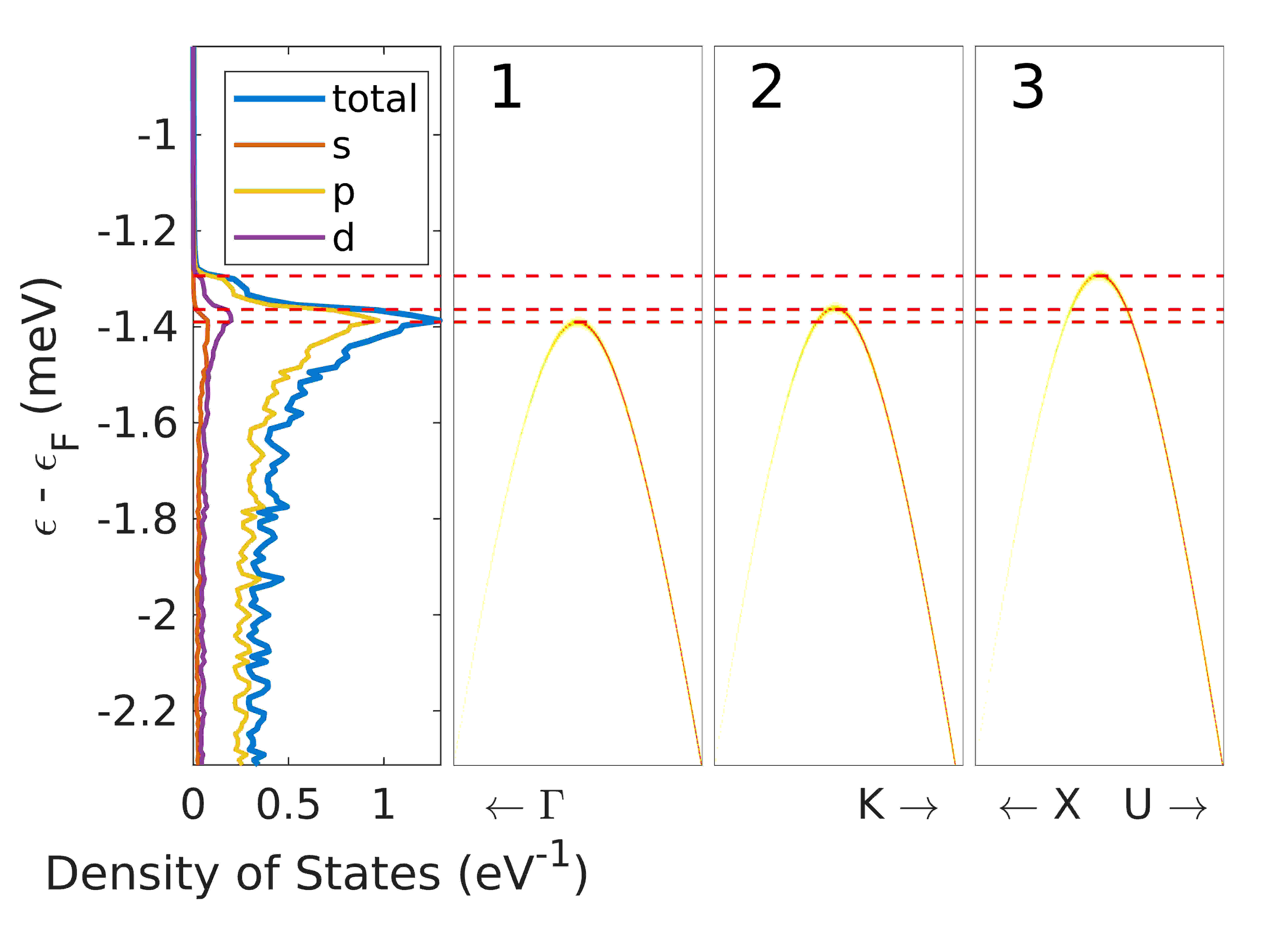}
\caption{Left Panel: The total and orbital resolved density of states of Pb near the Fermi level. Panels 1-3: Band gaps in the $\Gamma$ to $K$ and $X$ to $U$ directions. Panels 1 and 2 are associated with the larger gap in the DOS and panel 3 is associated with the smaller gap. The units of the spectral function are arbitrary.} 
\label{fig:GapAnisotropyPb}
\end{figure}

For an interaction parameter of $\Lambda=0.348 \mathrm{Ry}$ the over all gap size is found to be comparable to Ruby \textit{et al} \cite{Ruby2015b}. The inset of Fig.~\ref{fig:DOSPb} displays the gap structure at this interaction parameter where the separation between the distinct gaps is $80\mathrm{\mu eV}$. This energy separation is of comparable to the $150\mathrm{\mu eV}$ identified by Ruby \textit{et al} \cite{Ruby2015}. In order to resolve such small separation a fine mesh of $10 \times 10^6$ k-points in the irreducible part of the Brillouin zone was required. 

\begin{figure}[h]
\centering
\includegraphics*[width=1\linewidth,clip]{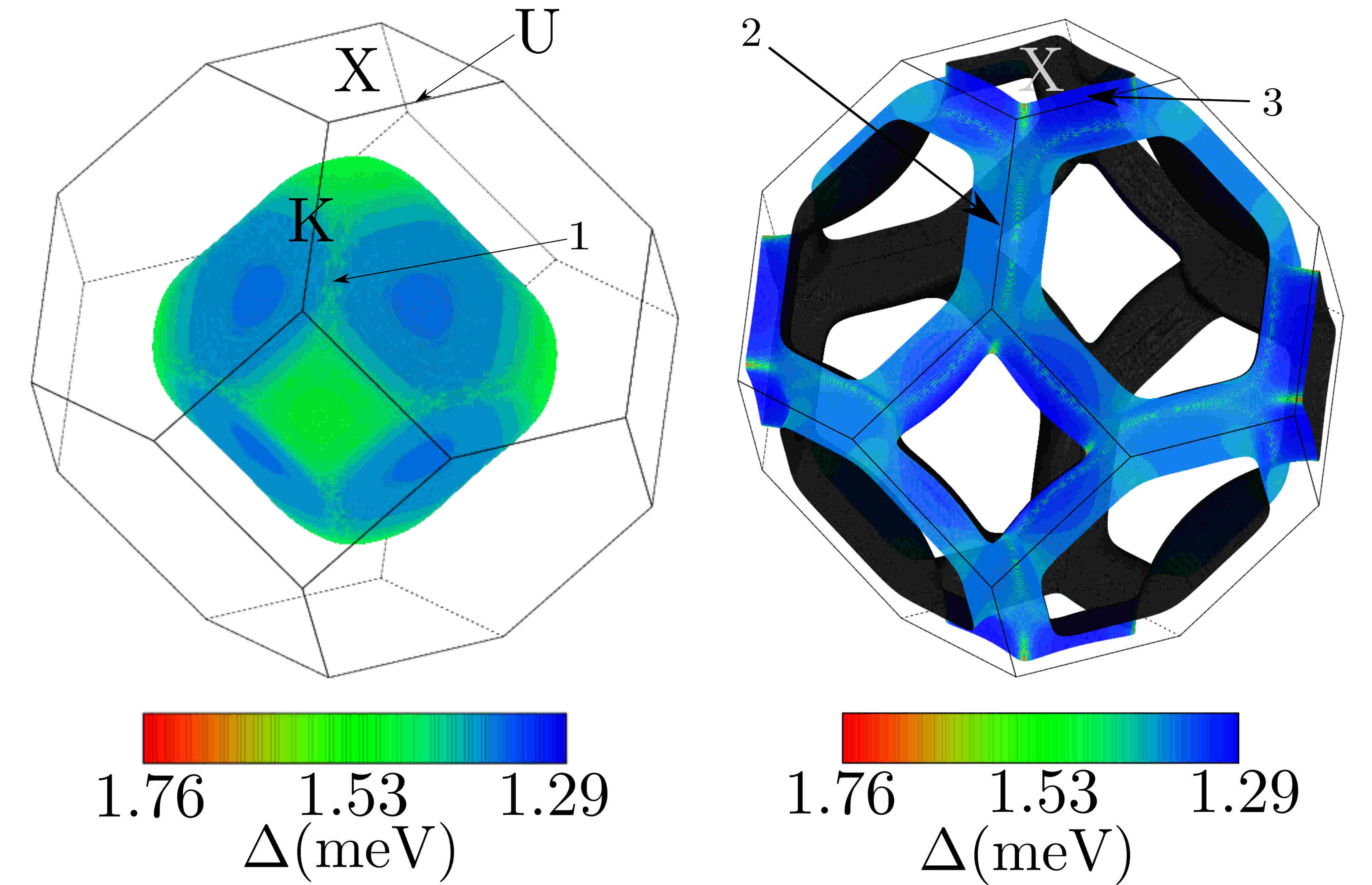}
\caption{The Fermi surface sheets of Pb in the normal state with the gap size of Pb in the superconducting state superimposed as a colour scale on top. The labels refer to the panels in Fig.~\ref{fig:GapAnisotropyPb} and identify the points on the Fermi surface where the gaps in Fig.~\ref{fig:GapAnisotropyPb} appear.} 
\label{fig:FermiPb_gap}
\end{figure}

Following the same process as for Nb the distinct gap sizes can be traced to different points in k-space as shown in Fig.~\ref{fig:GapAnisotropyPb}. However, to get the full picture of the anisotropy Fig.~\ref{fig:FermiPb_gap} shows the size of the gap on both Fermi surface sheets, with the relevant directions and band crossings from Fig.~\ref{fig:GapAnisotropyPb} highlighed. It is clear that the Fermi sheet in the left panel is mainly associated with the larger of the two gaps, and the sheet in the right panel contributes to the smaller shoulder. However, both sheets do contribute to a lesser extent to the other gaps as well. This is in contrast to the argument put forward by Ruby \textit{et al} \cite{Ruby2015b} who argued that the closed Fermi sheet (left panel of Fig.~\ref{fig:FermiPb_gap}) contributes to the smaller gap and the open sheet (right panel of Fig.~\ref{fig:FermiPb_gap}) relates to the outer peak. There are many factors which could be responsible for this discrepancy. Perhaps the most obvious is the fact that we performed a bulk calculation, whereas their experiment probes the surface states. An accurate calculation of the surface is therefore crucial to shed light on this aspect.

\begin{figure}[h]
\centering
\includegraphics*[width=1\linewidth,clip]{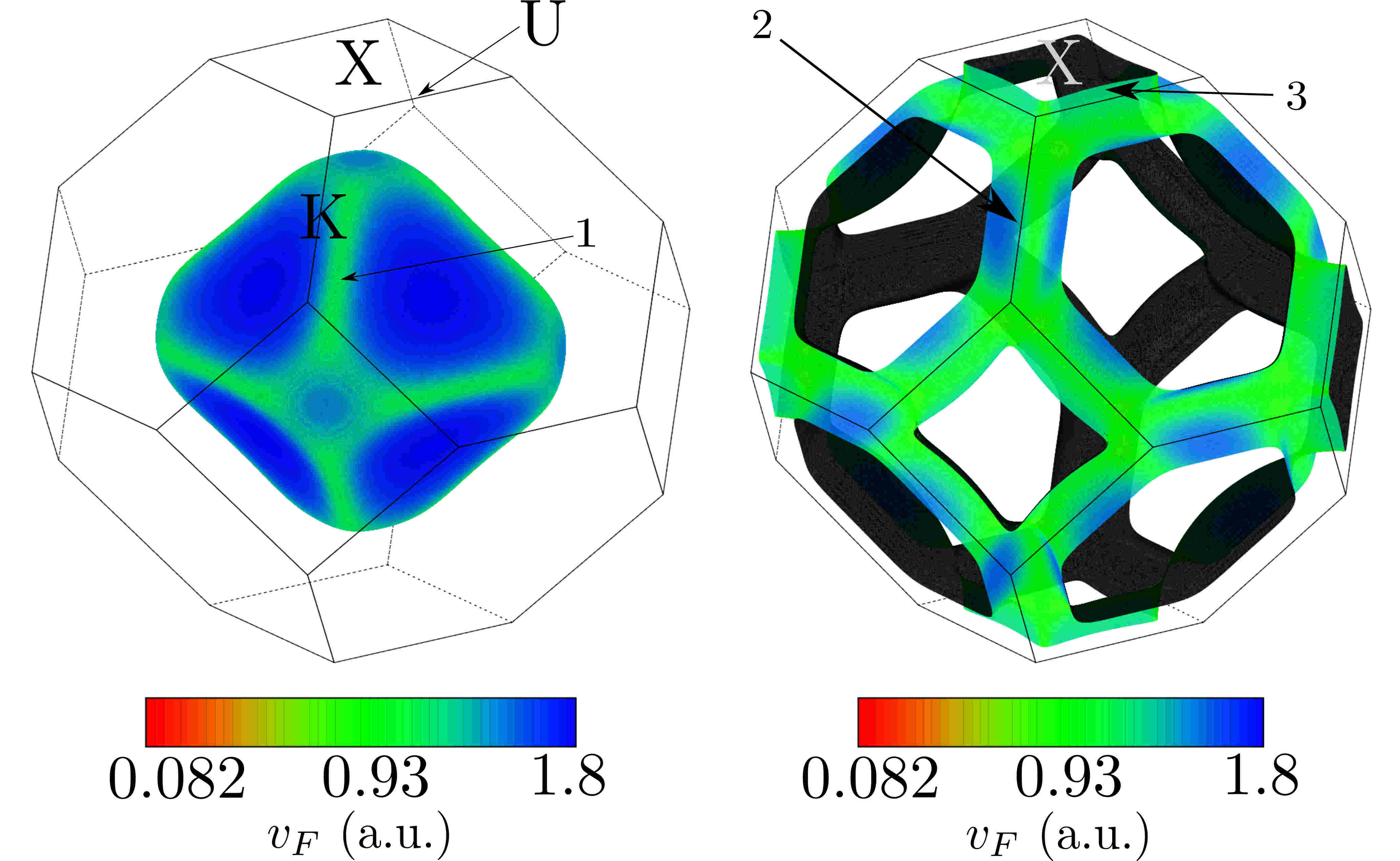}
\caption{The Fermi surfaces with the colour scale showing the Fermi velocity of Pb. The labels refer to the gaps identified in Fig.~\ref{fig:GapAnisotropyPb}.} 
\label{fig:FermiPb_vf}
\end{figure}

Interestingly, the Fermi velocities as shown in Fig.~\ref{fig:FermiPb_vf} together with the simple argument developed for Nb would suggest the open band to show the larger gap in agreement with the experimental observation. While for the closed Fermi surface sheet the simple argument connecting the Fermi velocity to the size of the gap largely holds the relationship is less convincing for the open sheet. This aspect deserves further investigation and in particular the influence of surfaces and spin-orbit coupling has to be addressed in future work. 

Other theoretical work on Pb by A. Floris \textit{at al.}\cite{Floris2007} established the anisotropic electron phonon coupling from fully first principles calculations for a bcc crystal structure. Since this is a different crystal structure, it complicates the comparison. The extracted degree of anisotropy, measuring the relative difference between the gaps on the two Fermi surface sheets, was $\approx 20\%$. For the fcc crystal considered here, the same anisotropy measure is $\approx 4\%$.


\subsection{\label{sec:4c} Magnesium Diboride}

\begin{figure}[h]
\centering
\includegraphics*[width=1\linewidth,clip]{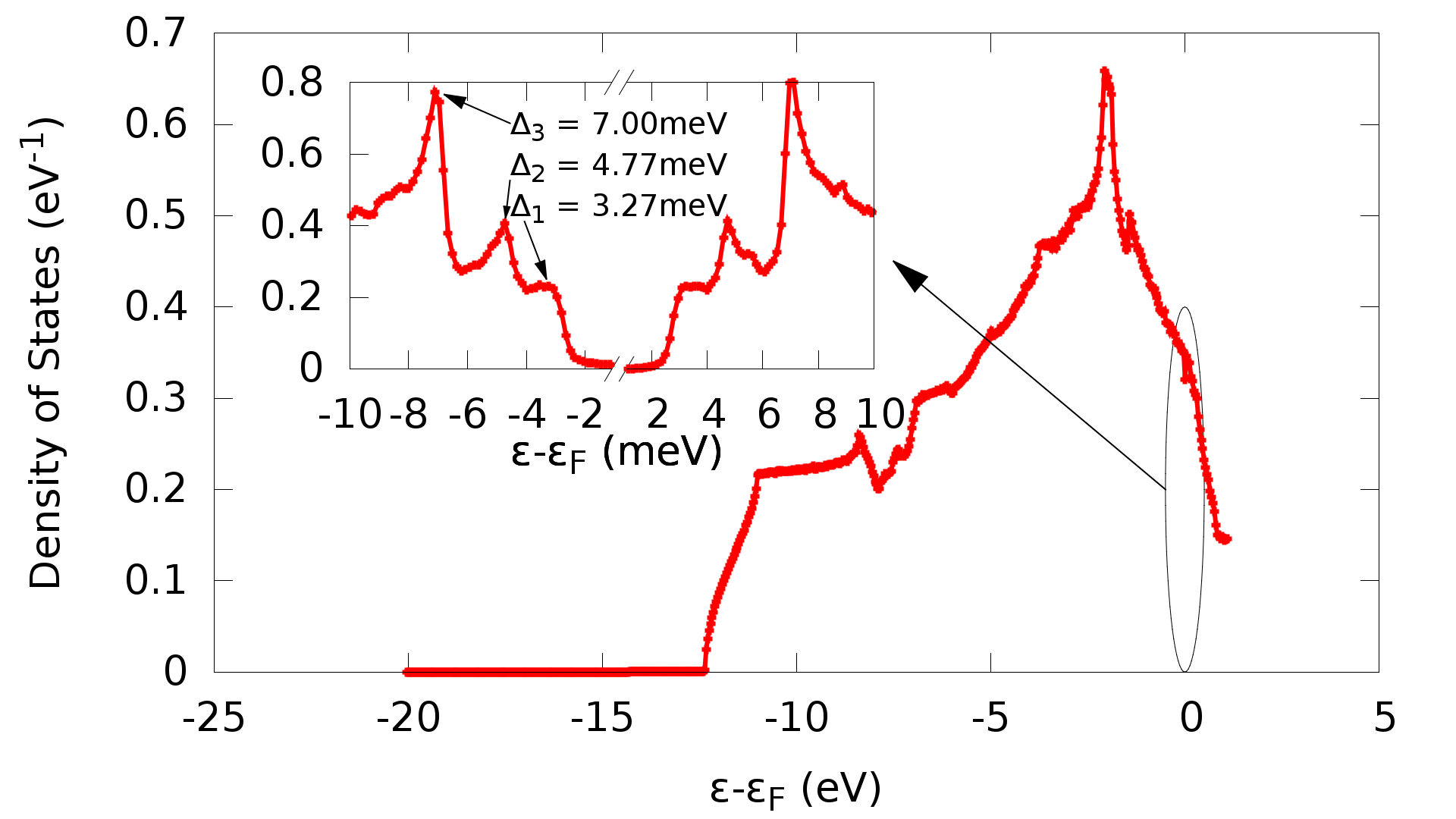}
\caption{The electronic density of states of MgB$_2$ in the superconducting state. Inset shows the density of states around the superconducting gap.} 
\label{fig:MgB2_supDOS}
\end{figure}

The third example is Magnesium diboride, which is well established as a phonon-mediated high temperature superconductor. Its superconductivity is mainly driven via a large E$_{2g}$ mode derived from the Boron atom~\cite{Budko2001,Mazin2003,Shukla2003}. For this reason, we model the superconducting state with the interaction on the Mg atoms $\Lambda_{Mg}=0$ and the interaction parameter for the Boron atoms as $\Lambda_{B}=0.288\mathrm{Ry}$. This parameter is tuned to fit the experimental zero temperature gap size~\cite{Choi2002,Szabo2001,Mou2015}. In experimental studies, the smaller gap at zero temperature ranges from $1.8\mathrm{meV}$ \cite{Choi2002} to approximately $3\mathrm{meV}$ \cite{Szabo2001,Mou2015}, whereas the larger gap is around $7\mathrm{meV}$. In Fig.~\ref{fig:MgB2_supDOS} we present the density of states within the superconducting state with the lattice constants $a=5.8317$ a.u. and $c=6.6594$ a.u.~. 


\begin{figure}[h]
\centering
\includegraphics*[width=1\linewidth,clip]{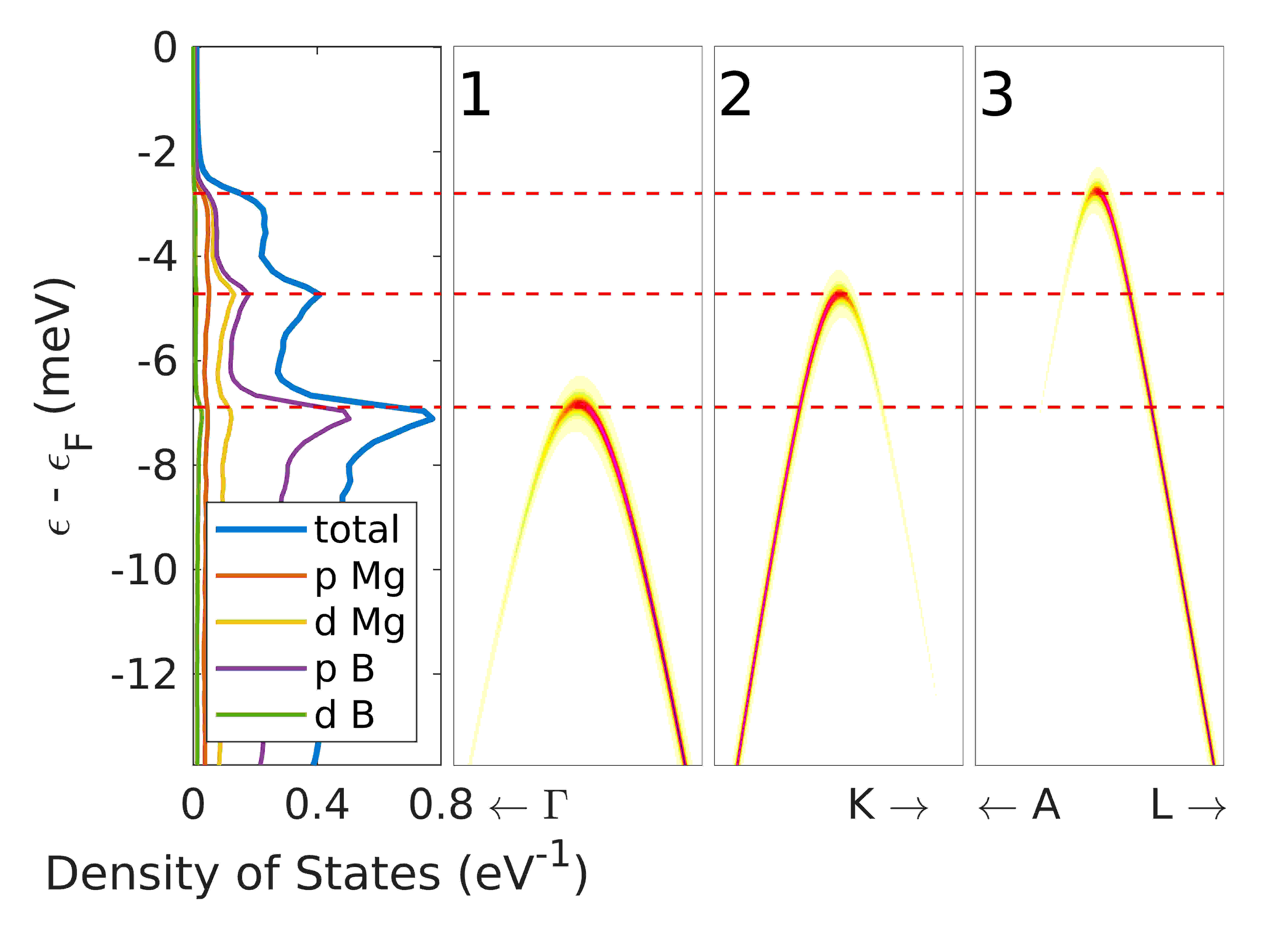}
\caption{
Left Panel: The total and atom orbital resolved density of states of MgB$_2$ near the Fermi level. Panels 1-3: Band gaps in the $\Gamma$ to $K$ and $A$ to $L$ directions. Panel 1 contributes to the largest gap in the DOS, panel 2 contributes to the middle gap and panel 3 is associated with the smallest gap. The units of the spectral function are arbitrary.} 
\label{fig:MgB2_gapAnisotropy}
\end{figure}

\begin{figure}[h]
\centering
\includegraphics*[width=1\linewidth,clip]{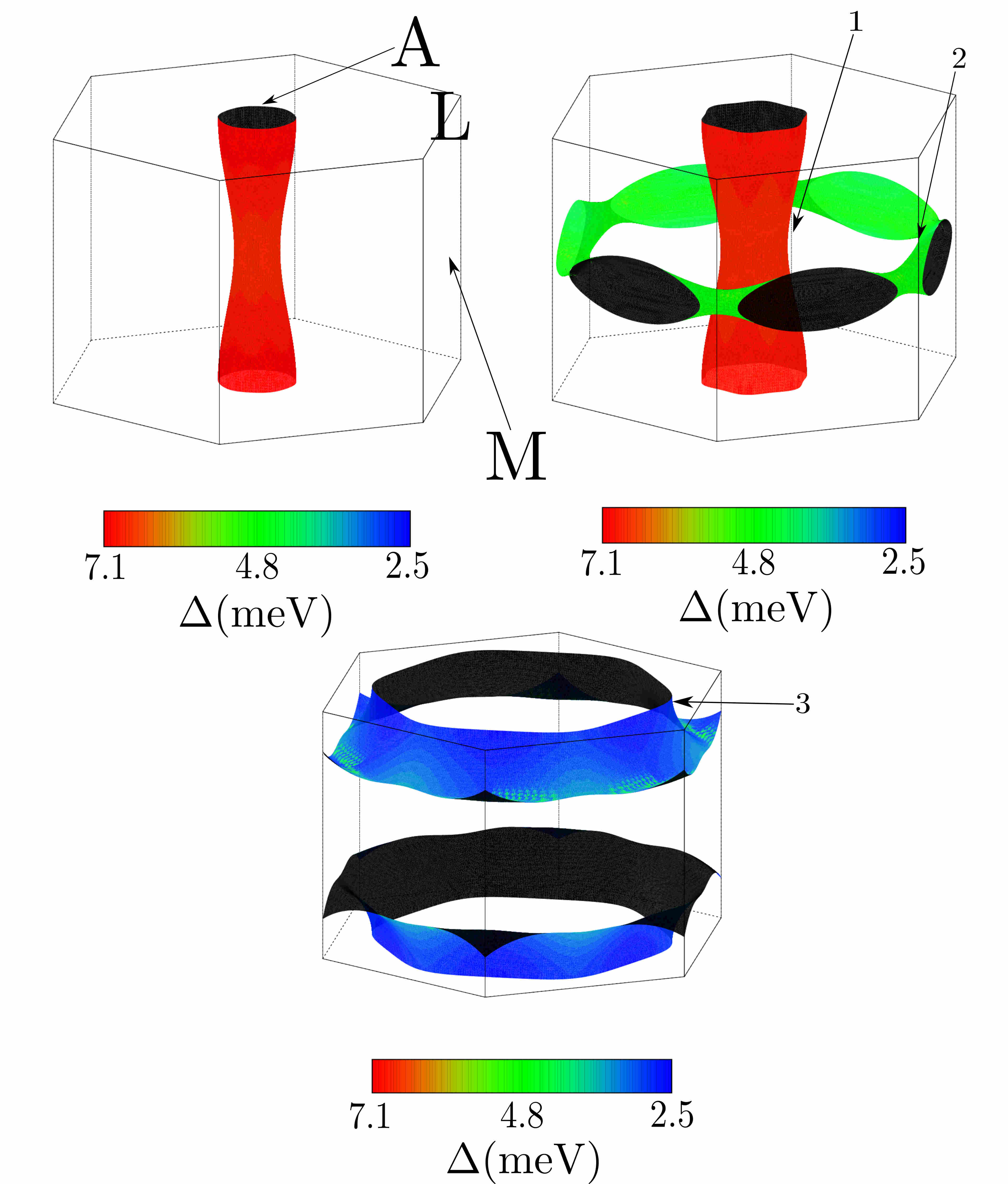}
\caption{The Fermi surfaces of MgB$_2$ in the normal state with the gap size in the superconducting state superimposed as a colour scale on top. The labels refer to the panels in Fig.~\ref{fig:MgB2_gapAnisotropy} and identify the points on the Fermi surface where the gaps in Fig.~\ref{fig:MgB2_gapAnisotropy} appear.} 
\label{fig:FermiMgB2_gap}
\end{figure}

There is very good agreement between the experimental gap sizes and the smallest gap at $\Delta_1=3.27\mathrm{meV}$ and the largest at $\Delta_3=7.00\mathrm{meV}$, which can be clearly identified in Fig.~\ref{fig:MgB2_supDOS}.  In addition, there is a third peak associated with a third superconducting gap. Fig.~\ref{fig:MgB2_gapAnisotropy} shows three band gaps in some high symmetry directions associated with each of the three peaks. As before  Fig.~\ref{fig:FermiMgB2_gap} extends this analysis to the  full Fermi surface. Each sheet can be associated with one of the three distinct gaps going from $\Delta_1=3.27\mathrm{meV}$ (blue) over $\Delta_2=4.77\mathrm{meV}$ (green) to $\Delta_3=7.00\mathrm{meV}$ (red). 

\begin{figure}[h]
\centering
\includegraphics*[width=1\linewidth,clip]{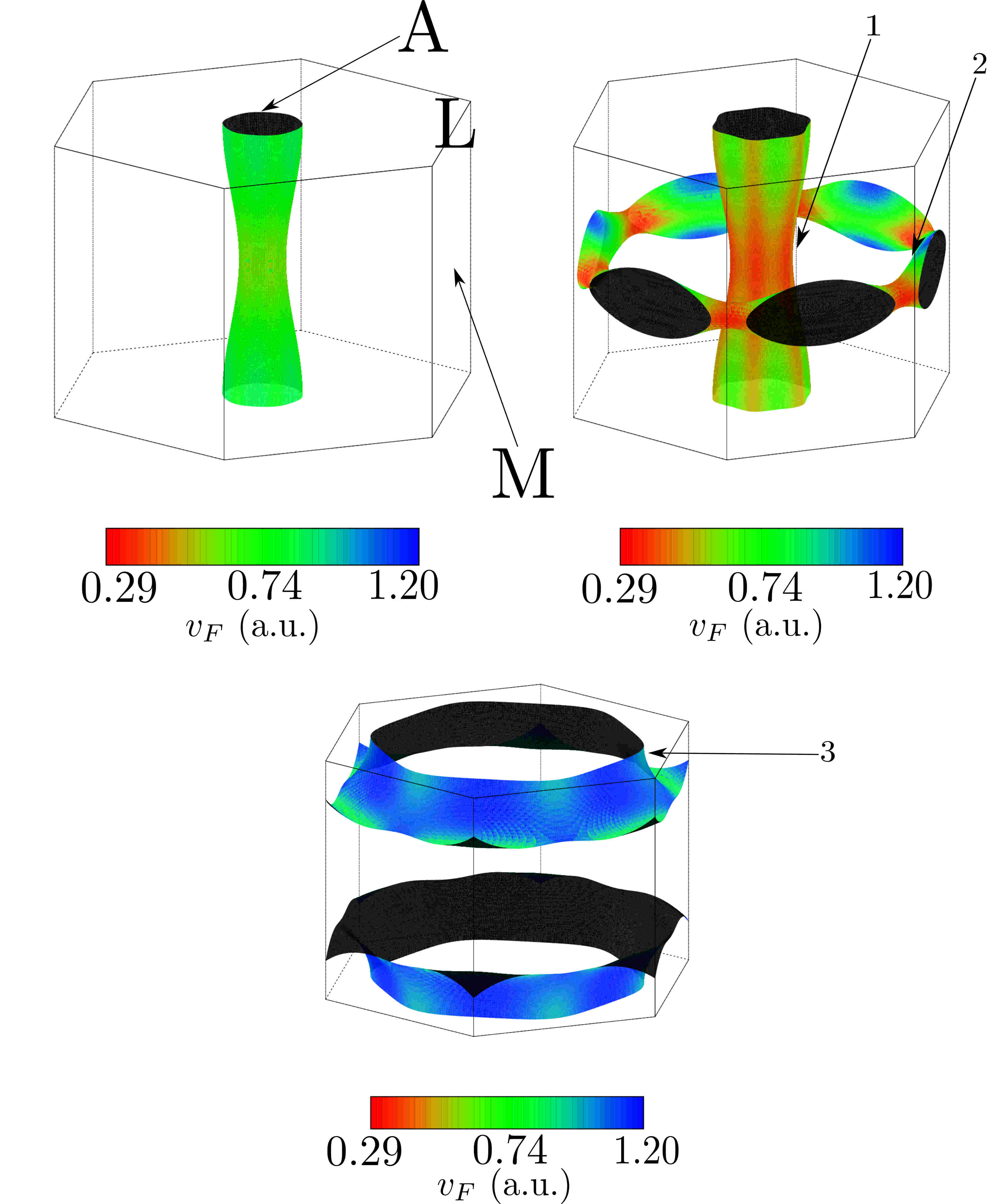}
\caption{The Fermi surfaces with the colour scale showing the Fermi velocity of MgB$_2$. The labels refer to the gaps identified in Fig.~\ref{fig:MgB2_gapAnisotropy}. } 
\label{fig:FermiMgB2_vf}
\end{figure}

Comparing this result to the Fermi velocities (Fig.~\ref{fig:FermiMgB2_vf}) the simple relation, as established for Nb, holds to a certain extent. However for the top right panel the Fermi velocities vary quite remarkably across the Fermi surface sheets whereas the gap is relatively constant across each of the individual sheets (see Fig.~\ref{fig:FermiMgB2_gap}).

The parameter-free calculation from A. Floris \textit{et al} \cite{Floris2005} accurately predicts both gaps at $\mathrm{T}=0\mathrm{K}$ and derives the correct transition temperature $T_c$ of MgB$_2$. This theoretical study, among others \cite{Margine2013,Mazin2003,Choi2002,Liu2001}, predicts two distinct gaps in the superconducting state. J. Bekaert \textit{et al} \cite{Bekaert2017} identifies a third gap for thin films of MgB$_2$. This third peak vanishes going beyond a thickness of 3~MLs highlighting the importance of out of plane hybridization. This suggests that any result will subtly depend on the out of plane lattice constant which we fixed to the experimental value rather using structural relaxation. Relaxed structures would show a smaller lattice constant possibly increasing out of plane hybridization and suppressing the third gap. On the other hand, in experiments impurity scattering will broaden any gap structures making it difficult to resolve a possible third peak.

\section{Discussions and Conclusions}

We have implemented a self-consistent solution of the BdG equations into the 3D bulk screened KKR formalism. In order to model realistic bulk systems, the BdG equation was solved self-consistently by choosing a simple exchange correlation functional (\ref{eqn:Suvasini}) to model both the normal and superconducting order parameters $\rho(\mathbf{r})$ and $\chi(\mathbf{r})$ respectively. The key parameter is the interaction $\Lambda$ which was tuned such that the zero temperature gap size matched the predicted experimental gaps. 

Calculating the gap self-consistently for Niobium we found two superconducting gaps from distinct Fermi surface sheets. We argued that the Fermi velocity of these bands at the Fermi level plays a key role in the indicated gap anisotropy.  The full anisotropic gap on the Fermi surface supported the simple picture connecting the gap size to the inverse of the Fermi velocity. The difference between the $\Delta_3$ and $\Delta_1$ is $0.36\mathrm{meV}$, and is comparable to experimental results \cite{Dobbs1964,MacVicar1968,Berndt1970,Klaumunzer1974}. The most recent tunnelling experiment gives roughly the same gap sizes~\cite{Hahn1998} which are $\Delta_{001} = 1.20$meV, $\Delta_{110} = 1.79$meV and $\Delta_{111} = 1.64$meV for the different planes. Similarly our calculations gave 3 distinct gaps of $\Delta_{1} = 1.43$meV, $\Delta_{2} = 1.69$meV and $\Delta_{3} = 1.79$meV. These gaps are very comparable to the tunneling experiment.

For fcc lead we identified two main gaps different by $80\mathrm{\mu eV}$. This is comparable to tunnelling experiments performed by Ruby \textit{et al} \cite{Ruby2015b} on single crystalline lead surfaces where they found a difference of $150\mathrm{\mu eV}$. However, our calculations do not support their Fermi-surface analysis. The future aim will be to investigate the real surfaces based on the method established here.

For MgB$_2$ we established gap sizes of similar order as found in the literature \cite{Szabo2001,Choi2002,Mou2015}. Within this framework we identified three superconducting gaps which is not unexpected in a system with three bands crossing the Fermi level and rather varied Fermi velocities. Three gaps have been predicted theoretically before in thin films only \cite{Bekaert2017}.

Experiments by Ruby \textit{et al} probed the influence of magnetic impurities placed on the surface of superconducting Pb \cite{Ruby2016,Ruby2018}. The presented  formalism here can be readily extended to such impurity systems, enabling the study of localised Yu-Shiba-Rusinov states\cite{Ruby2015,Ruby2016,Heinrich2018b}. Furthermore, implementing the fully relativistic BdG equations \cite{Csire2018} will include spin-orbit interaction. This, coupled with magnetic impurities, will make it possible to induce a triplet order parameter in these s-wave superconductors. 

In summary, we showed that using a fully \textit{ab initio} model to describe the normal state and a simple approximation for the superconducting exchange correlation functional produces gap anisotropy in Nb, Pb and MgB$_2$. This gap anisotropy is of the same order as experimental data for each of these systems. One of the key features of the formation of the gap anisotropy is the Fermi velocity at the Fermi surface which roughly correlates inversely with the magnitude of the gap. With just one free parameter, $\Lambda$, we have shown that it is possible to model anisotropy in superconductors to a high level of accuracy. In future it will be possible to extend this formalism to model superconductivity including impurities.  

\section{Acknowledgements}
The above work was supported by the Centre for Doctoral Training in Condensed Matter Physics, funded by EPSRC EP/L015544/1. B. \'Ujfalussy was supported by the Hungarian National Research, Development and Innovation Office under contract OTKA K115632. G. Csire gratefully acknowledges support from the European Union's Horizon 2020 research and innovation programme under the Marie-Sklodowska-Curie grant agreement No. 754510. The authors would like to thank M. Czerner and Prof Heiliger. In addition, thanks to Ming-Hung Wu and Reena Gupta for many helpful discussions. 

\bibliography{KKR_Superconductivity,KKR_Texts,General_Superconducting_Theories,SomeTextBooks,CES_Journal_Club2,CES_Journal_Club1,Pb,MgB2,STM_impurities,Yu-Shiba-Rusinov_States,Cuprate_Discovery,SC-DFT,BCS-BEC_papers,Sr2RuO4,BdGPhenomenological,MertigKKRCode,DestroyingSuperconductivity,AnisotropicNb,NoAnisotropyofNb,LDA+U}

\end{document}